# Complex hemodynamic responses to trans-vascular electrical stimulation of the renal nerve in anesthetized pigs

**Filippo Agnesi, PhD[1], Lucia Carlucci, PhD[1], Gia Burjanadze, MD[1], Fabio Bernini[1], Khatia Gabisonia, MD, PhD[1], John W Osborn, PhD[2] Silvestro Micera, PhD[3,4*], Fabio A. Recchia, MD, PhD[5,6,7*]**

1: Interdisciplinary Research Center "Health Science", Scuola Superiore Sant'Anna, Pisa, Italy
2: Department of Integrative Biology and Physiology, Medical School, University of Minnesota, Minneapolis, Minnesota, USA.
3: The BioRobotics Institute, Department of Excellence in Robotics & AI, Scuola Superiore Sant'Anna, Pisa, Italy
4: Bertarelli Foundation Chair in Translational NeuroEngineering, Centre for Neuroprosthetics and Institute of Bioengineering, École Polytechnique Fédérale de Lausanne (EPFL), Lausanne, Switzerland
5: Sector of medicine, Scuola Superiore Sant'Anna, Pisa, Italy
6: Lewis Katz School of Medicine, Cardiovascular Research Center, Temple University, Philadelphia, PA USA
7: Institute of Clinical Physiology, National Research Council, Italy.
*: These authors contributed equally to the manuscript.

**ABSTRACT** The objective of this study was to characterize hemodynamic changes during trans-vascular stimulation of the renal nerve and their dependence on stimulation parameters. We employed a stimulation catheter inserted in the right renal artery under fluoroscopic guidance, in pigs. Systolic, diastolic and pulse blood pressure and heart rate were recorded during stimulations delivered at different intravascular sites along the renal artery or while varying stimulation parameters (amplitude, frequency, and pulse width). Blood pressure changes during stimulation displayed a pattern more complex than previously described in literature, with a series of negative and positive peaks over the first two minutes, followed by a steady state elevation during the remainder of the stimulation. Pulse pressure and heart rate only showed transient responses, then they returned to baseline values despite constant stimulation. The amplitude of the evoked hemodynamic response was roughly linearly correlated with stimulation amplitude, frequency, and pulse width.

**INDEX TERMS** Electrical stimulation, Renal nerve, Trans-vascular stimulation.

**IMPACT STATEMENT** Hemodynamic responses to renal nerve trans-vascular electrical stimulation appears more complex than previously reported, fine combinations of stimulation parameters will be useful in clinical neuromodulation of the renal artery.

## I. Introduction

The kidneys are known to play a key role in the long-term arterial blood pressure (BP) regulation through two main mechanisms: production and release of renin and blood volume control by urinary sodium excretion. Renal innervation is of particular interest for its capacity to control renal function and to influence the activation state of the sympathetic system. In 1859 Claude Bernard observed that severing the greater splanchnic nerve produced ipsilateral diuresis and that electrical stimulation of the peripheral stub produced ipsilateral antidiuresis [1]. Nowadays, it is well known that sympathetic efferent fibres convey signals from the hypothalamic centers to the kidneys and afferent fibres originating from the kidneys transmit sympatho-excitatory signals towards autonomic regulatory nuclei in the central nervous system. The interplay between sympathetic nervous system and kidneys is thus bidirectional. Afferent and efferent fibres are intertwined to form the renal nerve that runs along the renal artery, partly attached to its adventitia. Sympathetic outflow regulates renal vascular resistance, renin release, and sodium re-absorption [2], [3], [4].

During the past decade, renal nerve radio-ablation through the renal artery has gained much attention as a potential therapy for the treatment of systemic hypertension resistant to pharmacological treatments. Reducing the sympathetic drive to kidneys is expected to lower peripheral vascular resistance while enhancing natriuresis. On the other hand, blocking the afferent signaling of the renal nerve would inhibit the renin-angiotensin-aldosterone system activation. Unfortunately, after initial positive results [5], [6], large clinical trials have not confirmed the efficacy of this approach [7] The factors that caused failure are still debated, but an important aspect that should be considered is the complexity of signaling traveling along the renal nerve. An alternative approach might consist of modulating the activity of the renal nerve, rather than ablating it, yet this strategy will require a deeper understanding of the nerve role in cardiovascular physiology. The classical approach to explore renal nerve function has been based on electrical stimulation, showing the effects of sympathetic renal nerve activity on renal and intrarenal hemodynamics, tubular solute, and water reabsorption, and on renin release3. While those studies have provided invaluable insights into the mechanisms of renal control, they were generally based on invasive experimental approaches which are not easy to translate to the clinical realm. Moreover, those observations were focused on renal, rather than cardiovascular responses. A few studies [8], [9], [10], [11], [12], [13], [14] employed intravascular electrodes to stimulate the renal nerve through the renal artery wall, showing an increase in BP during stimulation. Nevertheless, the description they provided was not extensive, as short periods of stimulation were used, the representation of the temporal dynamics of the systolic and diastolic BP changes as well as heart rate were limited (with one exception) to a single point at the end of the stimulation, and the effects of different stimulation parameters were not examined. This limits the usefulness of such data in understanding which physiological mechanisms are recruited during stimulation.

The aim of the present study was to test whether the hemodynamic response to trans-vascular electrical stimulation of the renal nerve displays heterogeneity, variations over time and dependence on stimulation parameters. Toward this goal we measured arterial systolic, diastolic, and pulse pressures and employed a post-processing approach capable of mitigating the oscillations related to the breathing cycle, thus allowing the visualization and characterization of small transient alterations.

## II. RESULTS

A total of 7 pigs were used during this study. The pattern of BP changes during electrical stimulation was complex, being characterized by negative and positive peaks preceding a phase of more stable augmentation (Fig. 1F and G). These features were not always present (i.e., some stimulations did not produce some or any of the initial peaks), whereas the "long term" response was sustained during stimulation in some cases, and in others a partial reduction was observed even before stopping the stimulation.

In 4 animals we applied the stimulation using fixed parameters (30 mA, 20 Hz, 5 ms PW) while moving the catheter along the renal artery by 2 mm steps. As shown in Fig. 2 A, B, C, and D, some degree of variability was observed among the hemodynamic responses to stimulations applied at different sites of the artery.

We sought to define the average pattern of the hemodynamic response and of its variability by superimposing the traces obtained from the stimulations delivered while moving the catheter (n=43 from 4 animals) and averaged them over time (thick black line in Fig. 2 E, F, G, and H). As shown in Fig. 2 E and F, systolic and diastolic BP responses to stimulation displayed an initial negative peak (N1) followed by a positive deflection (P1) and a second negative peak (N2). After these initial transient responses, BP values reached a maximum level (P2) followed by a partial reduction that was sustained until the stimulation was stopped. In contrast, the response pattern of PP and HR appeared simpler, with a single positive peak (named P2 for convenience, since it appears after the initial systolic and diastolic peaks). The time elapsed between the beginning of the stimulation and each peak is reported in Table I.

To characterize numerically these responses, we measured the changes in systolic and diastolic BP, PP and HR measured at the above-mentioned time points (n=43 from 4 animals). To better characterize the initial transient variations, P1 amplitude was calculated as the difference between the BP value at P1 and at N1, and N2 amplitude as the difference between the BP value at N2 and at P1. To evaluate the stability of the response during stimulation we also measured the amplitude of the "long term" phase by averaging the last 30 seconds before the end of the stimulation (EOS). As shown in Fig. 2 I, J, K, and L, changes from baseline were statistically significant for all measured time points and variables, apart from EOS for PP and HR Significance values are reported in Table I.

We then evaluated the dependency of P2 and EOS on stimulation intensity, frequency, and pulse width. In 4 animals we applied stimulations varying intensity, while keeping frequency and PW constant. Fig. 3 shows a relatively linear relationship between stimulation intensity and hemodynamic responses, both for P2 (Fig. 4 E, F, G, H) and EOS (Fig. 4, I, J, K, and L). We observed an almost complete absence of responses at the lowest amplitude.

In 4 animals we applied stimulations varying frequency, while keeping intensity and PW constant. In contrast to low intensities, hemodynamic changes were elicited even at lower stimulation frequencies, as shown in Fig. 4. Interestingly, whereas systolic and diastolic BP appears to increase linearly with the stimulation frequency (Fig. 5, E and F), a similar relationship did not occur for PP and HR (Fig. 4 G and H).

In 3 animals we varied the PW, while keeping intensity and frequency constant. Fig. 5 shows that small or no responses were observed at PW of 1 ms with a relatively linear increase of the evoked response to increasing levels of PW.

HRV was compared between during the stimulation and after the stimulation in n=43 recording from 4 animals. As can be seen in Table 2, SDNN and HRV were significantly elevated during stimulation when compared to the post-stimulation period. No significant difference was observed in the low to high frequencies ratio.

## III. DISCUSSION

The results of this study show the presence of a complex hemodynamic response to trans-vascular electrical stimulation of the renal nerve. Changes in BP were characterized by different phases, with negative and positive peaks. For example, the hemodynamic responses changed within the same animal when stimulation was applied at different sites along the renal artery.

To further characterize the hemodynamic response to trans-vascular electrical stimulation of the renal nerve, we tested the influence of stimulation parameters by altering one parameter at the time. We observed a progressive increase of the hemodynamic response mirroring the increase in stimulation amplitude, frequency, and PW. The almost complete absence of response at the lowest intensity value (5 mA) suggests that the electrical field generated by the stimulator needs to extend relatively far from the electrode to activate renal nerve fibres. This is in accordance with the available histologic data documenting that only a very small fraction of the fibres is located within the first half a millimeter of tissue surrounding the renal artery, both in swine [16] and in humans [17], [18]. The sympathetic fibres of the renal nerve are not organized in a single bundle, like in a somatic nerve, but are rather distributed within a relatively wide radius and some were found as far as 1 cm distant from the outer surface of the vessel wall [16]. Interestingly, the increase in hemodynamic responses seems to correlate almost linearly with the increase in the delivered current. This suggests that progressive recruitment of additional fibres occurs as the stimulation amplitude is increased. Moreover, the response amplitude did not reach a plateau, which may indicate that not all of the available fibres had been recruited even at the highest current intensity, pointing to the possibility that the electrical field generated in our experiments was not able to reach the nerve components furthest from the vessel wall.

The current intensity used during these stimulations was higher than expected and larger than those used in literature on canine models [8], [9], [10], [11], [13], albeit, comparable. While the distances between the nervous fibers and the renal artery is not very large, the presence of the arterial wall, as well as possible shunting effects related to the blood flow, appears to require significant intensities of current for recruiting renal nerve elements. Moreover, we utilized bipolar stimulation with the smallest inter-electrode distance (5 mm), which, differently from monopolar stimulation, is characterized by a much more limited spread of the current generated by the electrodes.

It is possible that trans-vascular stimulation methods that produce substantially wider activation fields would prove more efficient.

When specifically testing the impact of stimulation frequencies, we found a hemodynamic response even at the lower values (i.e., 5 and 10 Hz). Interestingly, whereas systolic and diastolic BP appeared to increase somewhat linearly in response to progressively higher stimulation frequencies, PP and HR responded differently. This is surprising, since, if the changes in PP and HR were entirely secondary to the systolic and diastolic BP alterations, one could expect them to follow the same pattern. Whereas such discrepancy could hint at the involvement of separate mechanisms, the limited number of data points prevents further interpretations.

Finally, we tested the specific impact of PW on the hemodynamic responses. In this case, all the measured hemodynamic parameters increased linearly in response to progressively longer PW. Since smaller axons have larger chronaxie values, increases in PW increase the number of recruited neurons activating fibres of progressively smaller axonal diameters. The need for such long PW values strongly suggests that most of the observed effects are likely mediated by small diameter unmyelinated fibres known to have a chronaxie around 2 ms [19]. Indeed, this is consistent with the fact that the sympathetic nervous system post-ganglionic neurons are C-fibres [20].

The difference in HRV between the stimulation and the post-stimulation period is not surprising considering the HR undergoes a transient increase as shown in Figure 2 (a). Any such alteration will inherently increase the standard deviation of the IBI as well as widen the IBI distribution as captured by the increase in SDNN and HRVI. Likewise, it is not surprising not to see significant changes in the LF/HF ratio. As the stimulation induced HR alteration occurs over a significant period, its contribution to the power spectrum is well below the range used to calculate the low frequency component of the LF/HF ratio.

A comparison with the existing literature is not straightforward, due to the variability in the stimulation parameters and duration utilized by other authors. In general, shorter stimulations, i.e., one [8], [9], [11], [13], [14] or two minutes [10] long, were used and often only single BP values recorded at the end of the stimulation. The previously reported increases in BP are larger than those observed in the current study and, interestingly, it was detectable 1 minute after the beginning of the stimulation, a time when, in our data, we instead observed the presence of N2. More in accordance with our results, others reported responses starting either "a few seconds" [9] or 15-30 secs [8] after the beginning of the stimulation. When a description of the time course was provided [14], the response appeared monotonic. It is hard to reconcile these discrepancies and they are likely related to differences in models (porcine versus canine), stimulation parameters, or anesthesia. Sevofluorane was used in our study compared to sodium thiopental [8], [13], sodium pentobarbital [11], [14], or isofluorane [9], [10]. Additionally, often, in our data post processing, the initial peaks became evident only after attenuating ventilation-related BP oscillations. Our results seem to confirm in pigs that BP is elevated during long term renal nerve trans-vascular stimulation, while we characterized more in depth the response time course.

There are several inherent study limitations. The presence of stimulation artifacts in the ECG recordings required us to estimate the IBI from the pressure signal. This approach is inherently less precise and, while it presents no problems when calculating HR, additional caution must be exercised with HRV as it could increase the inherent variability of the measurement. As bladder pressure was not monitored during stimulation, we cannot rule out that part of the observed effect could be due to a stimulation induced increase in bladder pressure influencing cardiac activity via afferent fibers. It is likely that the amplitude of the observed responses and the pattern of some of the time-dependent changes were affected by gas anesthesia. Other types of anesthesia utilized in the past for animal research are no longer an option. Additionally, unlike the clinical standard stimulating electrode employed by us, a specifically designed electrode should closely conform to the entire inner diameter of the vessel to ensure a uniform contact and circumferential current delivery. Finally, we did not compare hemodynamic changes in the absence and presence of pharmacological blockade of cardiac and vascular adrenergic response. Although desirable, these tests would have rendered difficult restoring baseline hemodynamic values before each new stimulation modality could be tested.

## IV. CONCLUSIONS

Our results provide important insights in the complexity of hemodynamic response to electrical renal nerve stimulation that did not emerge in previous studies and might contribute to the design of conservative and tunable therapeutic strategies as an alternative to renal nerve ablation.

## V. METHODS

### A. Ethical approval

The animal protocol was approved by institution's ethical committee (N. 1117) and by the Italian Ministry of Health and was in accordance with the Italian law (D.lgs. 26/2014). Experiments were carried out according to the guidelines laid down by the institution's animal welfare committee. We used 7 healthy male farm pig (Sus Scrofa) with a body weight ranging between 30 and 35 kg. They were pre-medicated with a cocktail of zolazepam (5mg/kg) and tiletamine (5 mg/kg), anesthetized with propofol (2 mg/kg intravenously), intubated and artificially ventilated. Anesthesia was maintained with 1-2 % sevoflurane vaporized in the respiratory 50% air-50% oxygen mixture. The animals were infused with 0.9 % NaCl solution over the duration of the experiment to prevent dehydration. A saline-filled catheter was inserted in the carotid artery to measure central arterial blood pressure, data were sampled at 1.5 kHz and saved for analysis. A 6Fr amplatz right (AR1.0) guide catheter (Medtronic, USA) was inserted in the femoral artery, advanced and positioned in the right renal artery under X-ray fluoroscopic guidance. Contrast medium was injected to confirm the correct positioning of the catheter (Fig 1A). Subsequently, a 4-cylindrical contacts stimulation catheter (Supreme, ,Abbott Chicago, IL, USA) was introduced in the renal artery to deliver electrical stimulations (Fig. 1B). This catheter has a 1.33 mm diameter and 4 electrodes, one covering the tip and 3 rings 2 mm long spaced 5 mm apart.

The stimulation catheter was connected to a constant current linear stimulator (Stmisola, Biopac, Goleta, CA, USA); set with output mode "I", z=100Ω, The stimulator was controlled using a digital to analog converter (National Instruments, Austin, Texas, USA). The DAQ was controlled using Matlab (Mathworks, Natik, MS, USA) using a sampling rate of 5000 Hz.

At the end of the study, animals were anesthetized with a bolus of propofol 1% (5 mg/Kg) and euthanized by injection of 10% KCl (20 ml) to stop the heart in diastole.

### B. Protocol

Stimulation was delivered using the most distal contact as the cathode and the immediately adjacent contact as the anode. The electrical stimuli consisted of cathodic first, biphasic symmetric rectangular waveforms with an intensity varying from 5 to 40 mA, a frequency between 5 and 50 Hz and a pulse width between 1 and 5 ms. In one set of experiments, stimulations were delivered at different points within the renal artery by moving the tip of the catheter from the proximal (near the aorta) to the distal end (toward the kidney) with 2 mm steps. During these experiments, the stimulation was delivered at 30 mA, 20 Hz and a pulse width (PW) of 5 ms. In a second set of experiments, the position of the catheter was fixed leaving it in the proximal segment of the artery, while varying intensity (5, 10, 20, 30, 40 mA, 20 Hz, 5 ms PW), frequency (5, 10, 20, 30, 50 Hz, 30 mA, 5 ms PW) or PW (1, 2, 3, 4, 5 ms, 30 mA, 20 Hz) of the stimuli.

During all the experiments, baseline hemodynamic values were recorded for 90 seconds before stimulation was delivered. Electrical stimulation was held for 5 minutes, followed by at least 5 minutes of pause to allow the return of BP to baseline levels (Fig 1C). Systoles and diastoles were detected for each heartbeat (Fig 1D). As expected, mechanical ventilation caused BP fluctuations synchronous with lung inflation and deflation (Fig. 1E). To reduce this confounding effect, systolic and diastolic BP were filtered using a mobile average filter with a window size corresponding to the number of heart beats recorded during each cycle of the respirator at baseline (Fig. 1F). To ensure consistency in the timing of each heartbeat detection, the BP signal was low passed with a 4th order Butterworth filter and a cut-off frequency of 5 Hz. Systolic peaks in this smoothed BP trace were identified and instantaneous HR was derived based on the inverse of the peak-to-peak time interval. As the respiratory cycle also affected HR, the mobile average filter used for systolic and diastolic BP was applied to the HR to minimize the impact of its oscillations. To allow a better comparison between different stimulations, systolic and diastolic BP and HR traces were normalized subtracting the average values calculated over the baseline phase (Fig. 1G and H). Pulse pressure (PP) was also calculated as the difference between systolic and diastolic BP.

To provide a more comprehensive picture, we analyzed heart rate variability (HRV), albeit with mayor limitations. Stimulation artifacts interfered significantly with ECG recordings and could not be used to identify R-R intervals. The beginning of the sharp rise in BP associated with each heartbeat was instead used as proxy to calculate inter-beat intervals (IBIs). IBIs were used to calculate the standard deviation of the IBI (SDNN), the heart rate variability index (HRVI) and the ratio between the power in the low and the high frequencies in the IBI spectrum (LF/HF) [15]. As the baseline period used for the acquisitions was inferior to the 5 minutes suggested for HRV calculations, the stimulation period was compared only to the post-stimulation period. HRV was calculated on the data obtained while moving the electrode position as they better represent the variability of the responses to trans-vascular renal nerve stimulation.

### C. Statistical Analysis

Data are reported as mean ± standard deviation. The average of peak changes in BP and HR were normalized to baseline values and relative changes vs 0 (null hypothesis) were tested using the one sample t test. To evaluate the dependency of the hemodynamic responses on stimulation parameters, we calculated the corresponding Pearson's coefficients and the associated significance values. Changes in HRV measurements between the stimulation period and the post stimulation period were tested using paired two sample t test.

| Peak | N1 | P1 | N2 | P2 | EOS |
|---|---|---|---|---|---|
| Systolic BP | | | | | |
| Delay (s) | 18.9 | 40.1 | 51.1 | 118.6 | NA |
| p-Val | 1.22E-06 | 4.46E-10 | 5.90E-04 | 5.15E-13 | 8.20E-11 |
| Diastolic BP | | | | | |
| Delay (s) | 16 | 35.1 | 56.5 | 124.1 | NA |
| p-Val | 4.19E-05 | 1.44E-07 | 1.45E-04 | 6.01E-10 | 1.34E-11 |
| Differential BP | | | | | |
| Delay (s) | NA | NA | NA | 96.2 | NA |
| p-Val | | | | 1.11E-06 | 0.131 |
| HR | | | | | |
| Delay (s) | NA | NA | NA | 64.1 | NA |
| p-Val | | | | 3.38E-07 | 0.244 |

**Table I:** Time elapsed from the beginning of stimulation and significance for each of the identified peaks (n=43 from 4 animals).

|  | ON | POST | p-Val |
|---|---|---|---|
| SDNN | 21.88±10.58 | 14.19±6.81 | 2.48E-05 |
| HRVI | 4.41±1.56 | 3.5±0.99 | 2.22E-05 |
| LF/HF | 0.89±0.49 | 0.87±0.43 | 0.59 |

Table 2: Mean ± standard deviation of the heart rate variability outcomes.

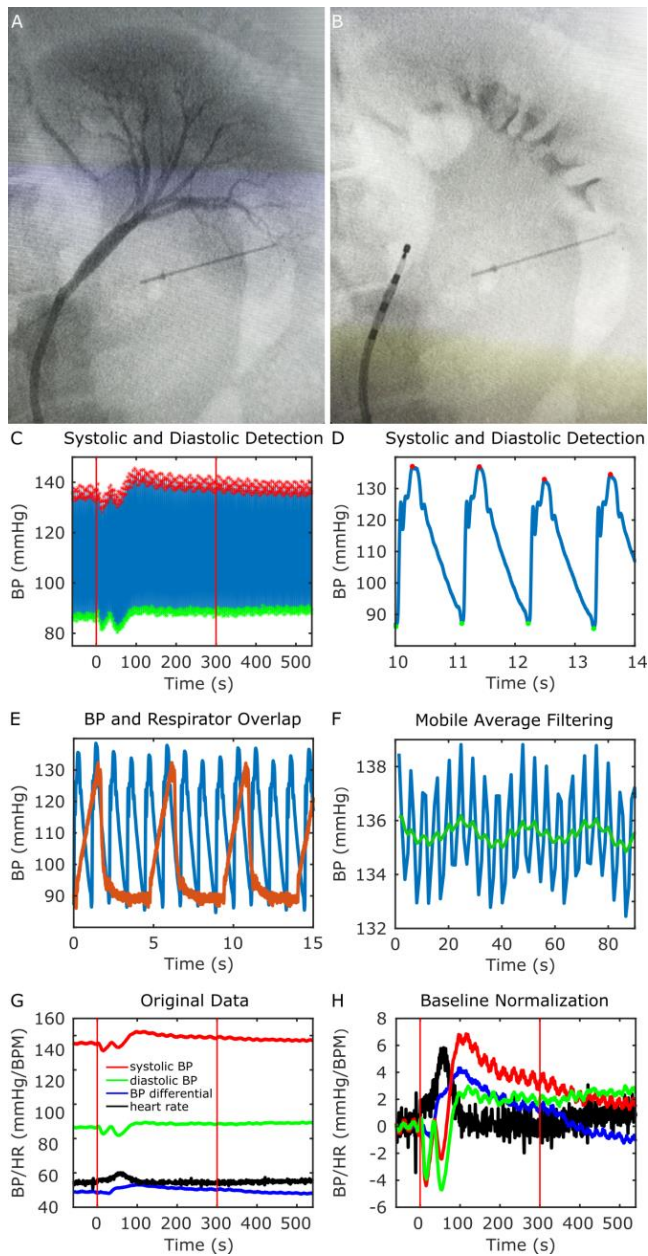

**Figure 1: experimental setup and hemodynamics recording and processing**. A: Fluoroscopy image of contrast medium diffusing in the renal artery. The straight line visible in the lower right side is a needle taped to the animal abdomen to provide a visual reference. B: Fluoroscopy image of the stimulation catheter advancing in the renal artery. C: Detection of systolic and diastolic BP values across an entire recording period. Vertical red lines represents the beginning and the end of the stimulation. D: Systolic and diastolic BP waveforms recorded for a few heartbeats. E: Superimposition of BP and respirator pressure waveforms. F: Reduction of oscillations related to the respiratory cycle obtained with mobile average filtering. G: Superimposition of systolic and diastolic BP, PP and HR. H: Superimposition of systolic and diastolic BP, pulse pressure and heart rate after baseline normalization.

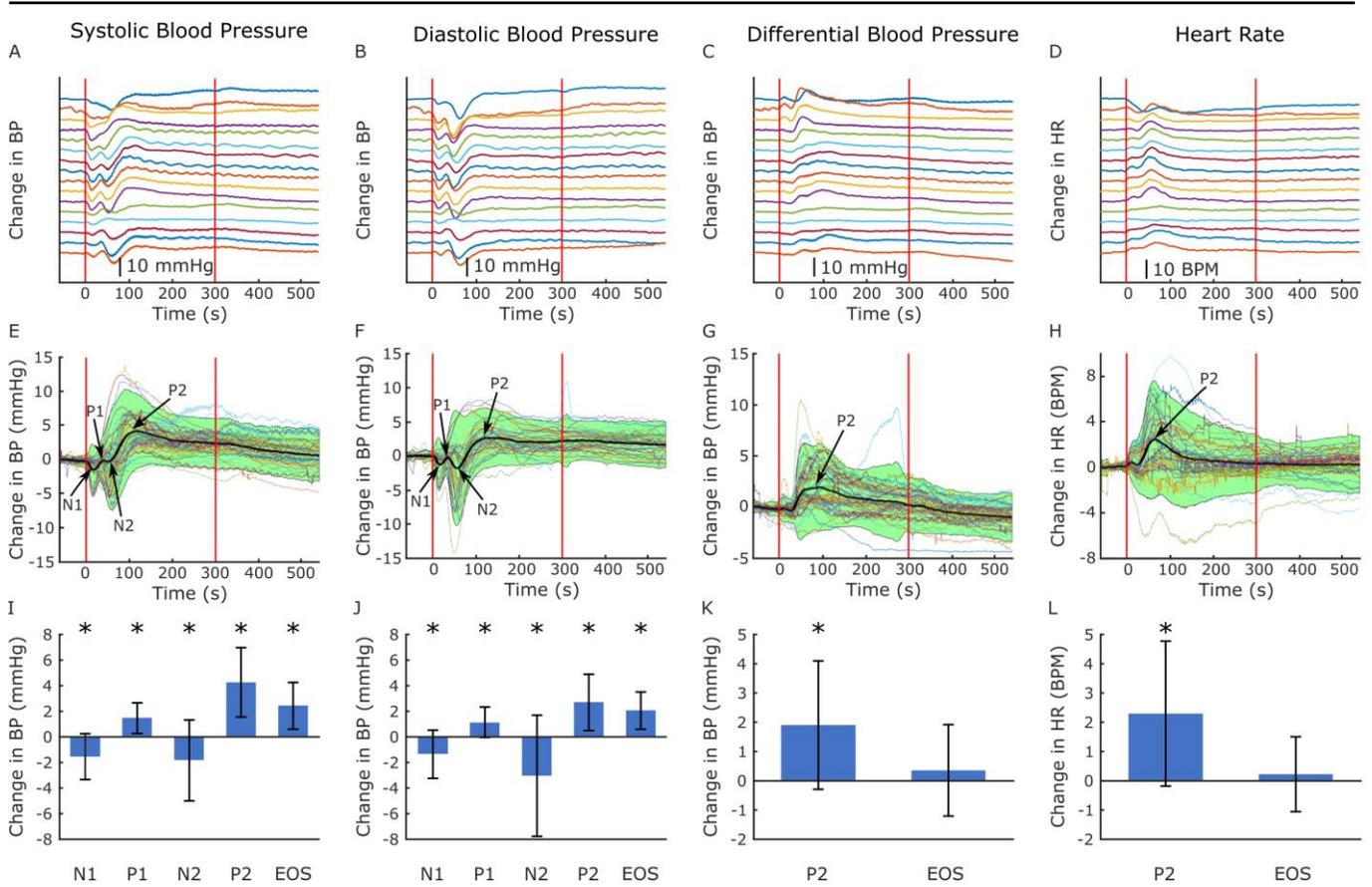

**Figure 2: Hemodynamic response variability.** A, B, C, and D: changes in systolic and diastolic BP, PP and HR elicited by stimulations delivered at different points along the renal artery. Each curve corresponds to a specific point, from the most proximal (top curve) to the most distal (bottom). The electrode was advanced from point to point by 2 mm steps. Red vertical lines represent the beginning and the end of the stimulation. E, F, G, and H: Superimposition of 43 responses obtained in 4 animals. Stimulations were delivered at 30 mA, 20 Hz and 5 ms PW. The black curve represents the average change. The shaded area ± two SD. I, J, K, and L: average ± standard error of the response magnitude for the different peaks. Asterisks indicate statistical difference from zero. P values are reported in table 1 (n=43 from 4 animals). EOS represents the mean value during the last 30 seconds of stimulation.

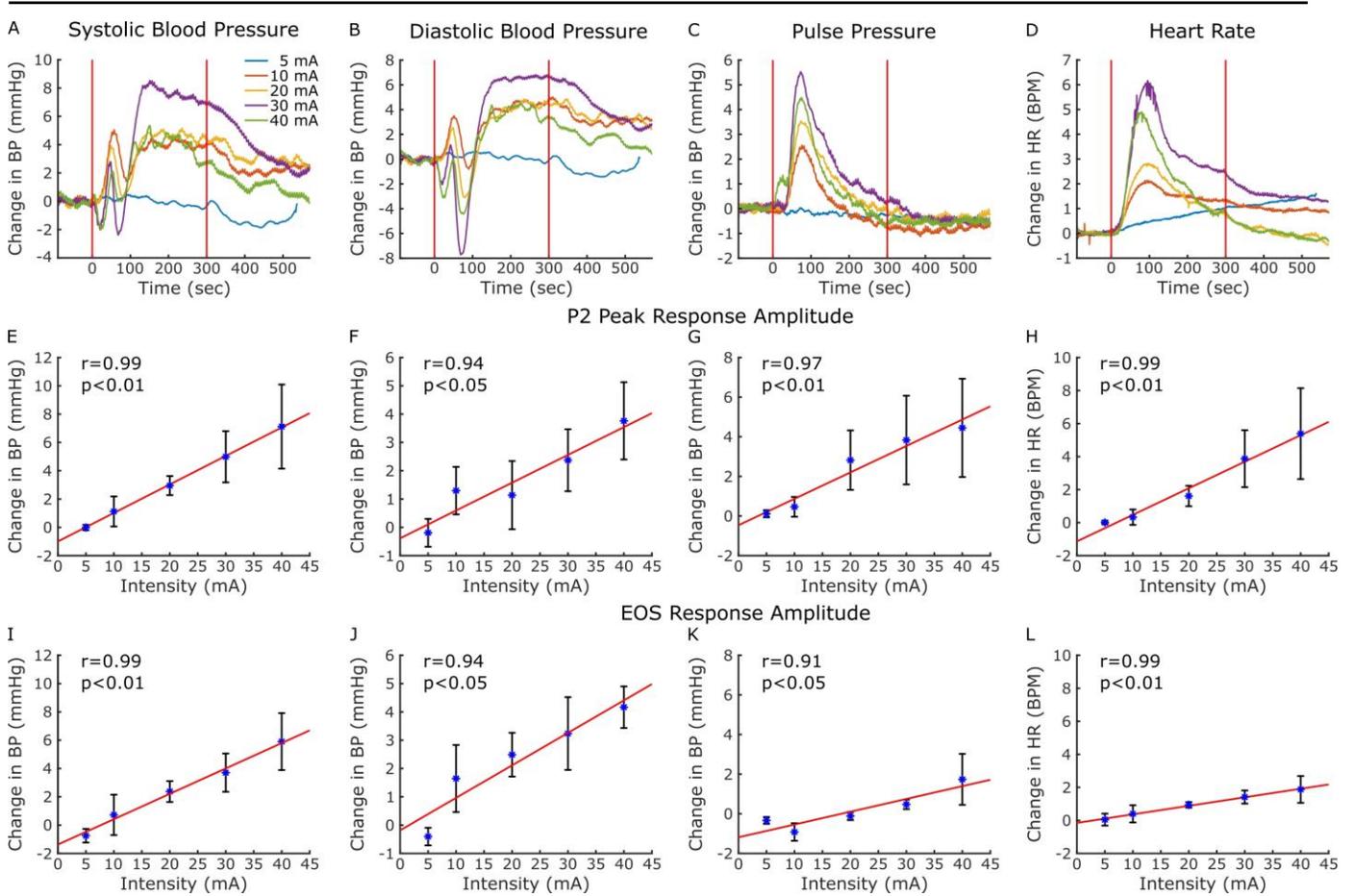

**Figure 3: Hemodynamic responses to varying stimulations intensities**. A, B, C, D: examples of responses to stimulations of different intensity in one animal. Red vertical lines represent the beginning and the end of the stimulation. E, F, G, H: average ± standard error of the magnitude of P2 evoked by different stimulation intensities. I, J, K, L: average ± standard error of the magnitude of the response at the end of stimulation (EOS) at different intensities. Pearson coefficient and relative significance level are reported for each correlation. (n=4 from 4 animals)

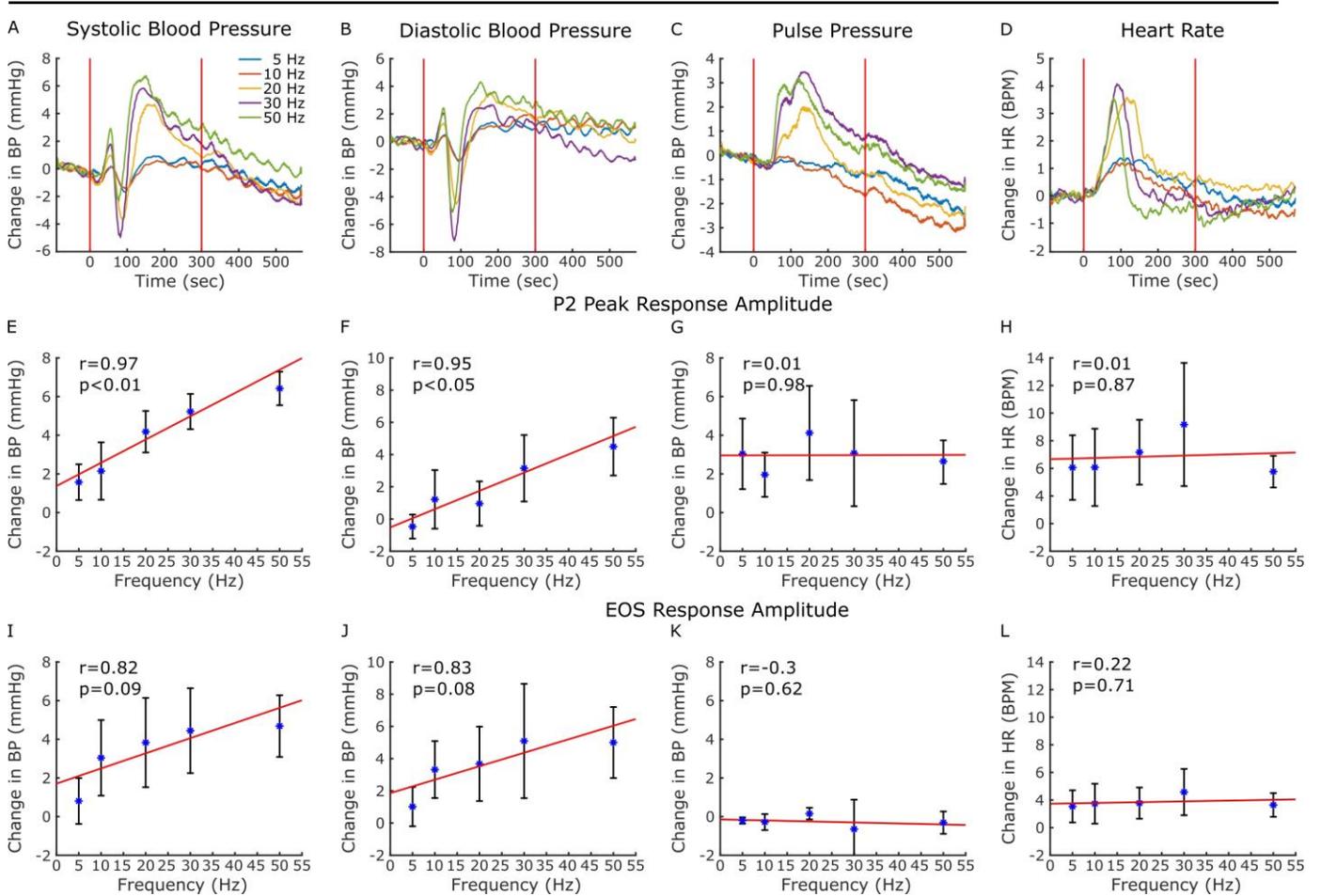

**Figure 4: Hemodynamic responses to varying stimulation frequencies**. A, B, C, D: example of the responses to stimulations at different frequencies in one animal. E, F, G, H: average ± standard error of the magnitude of P2 evoked by different stimulation frequencies. I, J, K, L: average ± standard error of the magnitude of the response at the end of stimulation (EOS) at different frequencies. Pearson coefficient and relative significance level are reported for each correlation. (n=4 from 4 animals)

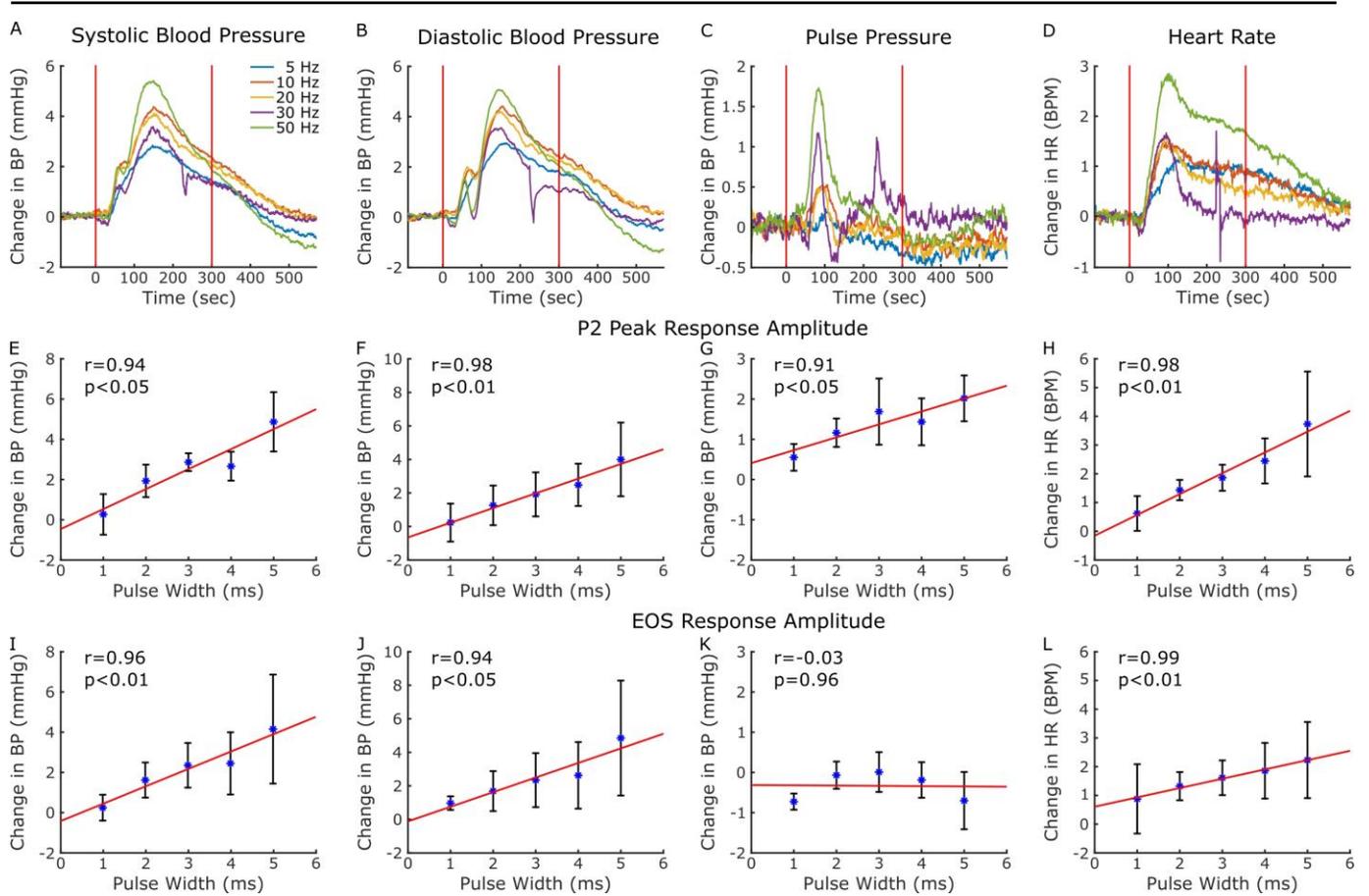

Figure 5: **Hemodynamic responses to varying stimulation pulse widths**. A, B, C, D: example of the responses to stimulations at different pulse widths in one animal. E, F, G, H: average ± standard error of the magnitude of P2 evoked by different stimulation pulse widths. I, J, K, L: average ± standard error of the magnitude of the response at the end of stimulation (EOS) at different pulse widths. Pearson coefficient and relative significance level are reported for each correlation. (n=3 from 3 animals)